\documentclass{article}

\begin{document}

\vspace{1cm}
\begin{flushright}
CAMS/00-04\\
\end{flushright}
\vspace{1cm} \baselineskip=16pt

\begin{center}
\baselineskip=16pt
\centerline{{\Large{\bf Complexified Gravity in
Noncommutative Spaces}}} \vskip1 cm

Ali H. Chamseddine \vskip1cm
\centerline{\em Center for Advanced
Mathematical Sciences (CAMS) } \centerline{\em and}
\centerline{\em Physics Department, American University of Beirut,
Lebanon}
\end{center}

\vskip1 cm \centerline{\bf ABSTRACT} The presence of a constant background
antisymmetric tensor for open strings or D-branes forces the space-time
coordinates to be noncommutative. This effect is equivalent to replacing
ordinary products in the effective theory by the deformed star product. An
immediate consequence of this is that all fields get complexified. The only
possible noncommutative Yang-Mills theory is the one with $U(N)$ gauge
symmetry. By applying this idea to gravity one discovers that the metric
becomes complex. We show in this article that this procedure is completely
consistent and one can obtain complexified gravity by gauging the symmetry $%
U(1,D-1)$ instead of the usual $SO(1,D-1)$. The final theory depends on a
Hermitian tensor containing both the symmetric metric and antisymmetric
tensor. In contrast to other theories of nonsymmetric gravity the action is
both unique and gauge invariant. The results are then generalized to
noncommutative spaces. \vfill\eject
\bigskip

\section{Introduction}

The developments in the last two years have shown that the presence of a
constant background $B$-field for open strings or D-branes lead to the
noncommutativity of space-time coordinates (\cite{CDS},\cite{DH},\cite{CK},
\cite{CH},\cite{S},\cite{AFS},\cite{SW}). This can be equivalently realized
by deforming the algebra of functions on the classical world volume. The
operator product expansion for vertex operators is identified with the star
(Moyal) product of functions on noncommutative spaces (\cite{H},\cite{FFZ}).
In this respect it was shown that noncommutative U(N) Yang-Mills theory does
arise in string theory.

The effective action in presence of a constant $B$-field background is
\[
-\frac{1}{4}\int Tr\left( F_{\mu \nu }\ast F^{\mu \nu }\right)
\]
where
\[
F_{\mu \nu }=\partial _{\mu }A_{\nu }-\partial _{\nu }A_{\mu }+iA_{\mu }\ast
A_{\nu }-iA_{\nu }\ast A_{\mu }
\]
and the star product is defined by
\[
f\left( x\right) \ast g\left( x\right) =e^{\frac{i}{2}\theta ^{\mu \nu }%
\frac{\partial }{\partial \zeta ^{\mu }}\frac{\partial }{\partial \eta ^{\nu
}}}f\left( x+\zeta \right) g\left( x+\eta \right) \left| _{\zeta =\eta
=0}\right.
\]
This definition forces the gauge fields to become complex. Indeed the
noncommutative Yang-Mills action is invariant under the gauge
transformations
\[
A_{\mu }^{g}=g\ast A_{\mu }\ast g_{\ast }^{-1}-\partial _{\mu }g\ast g_{\ast
}^{-1}
\]
where $g_{\ast }^{-1}$is the inverse of $g$ with respect to the star
product:
\[
g\ast g_{\ast }^{-1}=g_{\ast }^{-1}\ast g=1
\]
The contributions of the terms $i\theta ^{\mu \nu }$ in the star product
forces the gauge fields to be complex. Only conditions such as $A_{\mu
}^{\dagger }=-A_{\mu }$ could be preserved under gauge transformations
provided that $g$ is unitary: $g^{\dagger }\ast g=g\ast g^{\dagger }=1.$ It
is not possible to restrict $A_{\mu }$ to be real or imaginary to get the
orthogonal or symplectic gauge groups as these properties are not preserved
by the star product (\cite{SW},\cite{MW}). I\ will address the question of
how is gravity modified in the low-energy effective theory of open strings
in the presence of background fields. It has been shown that the metric of
the target space gets modified by contributions of the $B$-field and that it
becomes nonsymmetric (\cite{CLNY},\cite{SW}). If we think of gravity as
resulting from local gauge invariance under Lorentz transformations in the
tangent manifold, then the previous reasoning would suggest that the
vielbein and spin connection both get complexified with the star product.
This seems inevitable as the star product appears in the operator product
expansion of the string vertex operators.

We are therefore led to investigate whether gravity in D dimensions can be
constructed by gauging the unitary group $U(1,D-1)$. In this article we
shall show that this is indeed possible and that one can construct a
Hermitian action which governs the dynamics of a nonsymmetric complex
metric. Once this is achieved, it is straightforward to give the necessary
modifications to make the action noncommutative. The plan of this paper is
as follows. In section two the action for nonsymmetric gravity based on
gauging the group $U(1,D-1)$ is given and the structure of the theory
studied. In section three the equations of motion are solved to make
connection with the second order formalism. In section four we give the
generalization to noncommutative spaces. Section five is the conclusion.

\section{Nonsymmetric gravity by gauging U(1,D-1)}

Assume that we start with the $U(1,D-1)$ gauge fields $\omega _{\mu
\,\,b}^{\,a}$. The $U(1,D-1)$ group of transformations is defined as the set
of matrix transformations leaving the quadratic form
\[
\left( Z^{a}\right) ^{\dagger }\eta _{b}^{a}Z^{b}
\]
invariant, where $Z^{a}$ are $D$ complex fields and
\[
\eta _{b}^{a}=diag\left( -1,1,\cdots ,1\right)
\]
with $D-1$ positive entries. The gauge fields $\omega _{\mu \,\,b}^{\,a}$
must then satisfy the condition
\[
\left( \omega _{\mu \,\,b}^{\,a}\right) ^{\dagger }=-\eta _{c}^{b}\omega
_{\mu \,\,d}^{\,c}\eta _{a}^{d}
\]
The curvature associated with this gauge field is
\[
R_{\mu \nu \,\,b}^{\quad a}=\partial _{\mu }\omega _{\nu
\,\,b}^{\,a}-\partial _{\nu }\omega _{\mu \,\,b}^{\,a}+\omega _{\mu
\,\,c}^{\,a}\omega _{\nu \,\,b}^{\,c}-\omega _{\nu \,\,c}^{\,a}\omega _{\mu
\,\,b}^{\,c}
\]
Under gauge transformations we have
\[
\widetilde{\omega }_{\mu \,\,b}^{\,a}=M_{c}^{a}\omega _{\mu
\,\,d}^{\,c}M_{b}^{-1d}-M_{c}^{a}\partial _{\mu }M_{b}^{-1c}
\]
where the matrices $M$ are subject to the condition:
\[
\left( M_{c}^{a}\right) ^{\dagger }\eta _{b}^{a}M_{d}^{b}=\eta _{d}^{c}
\]
The curvature then transforms as
\[
\widetilde{R}_{\mu \nu \,\,b}^{\quad a}=M_{c}^{a}R_{\mu \nu \,\,d}^{\quad
c}M_{b}^{-1d}
\]
Next we introduce the complex vielbein $e_{\mu }^{a}$ and its inverse $%
e_{a}^{\mu }$ defined by
\begin{eqnarray*}
e_{a}^{\nu }e_{\mu }^{a} &=&\delta _{\mu }^{\nu } \\
e_{\nu }^{a}e_{b}^{\nu } &=&\delta _{b}^{a}
\end{eqnarray*}
which transform as
\begin{eqnarray*}
\widetilde{e}_{\mu }^{a} &=&M_{b}^{a}e_{\mu }^{b} \\
\widetilde{e}_{a}^{\mu } &=&\widetilde{e}_{b}^{\mu }M_{a}^{-1b}
\end{eqnarray*}
It is also useful to define the complex conjugates
\begin{eqnarray*}
e_{\mu a} &\equiv &\left( e_{\mu }^{a}\right) ^{\dagger } \\
e^{\mu a} &\equiv &\left( e_{a}^{\mu }\right) ^{\dagger }
\end{eqnarray*}
With this, it is not difficult to see that
\[
e_{a}^{\mu }R_{\mu \nu \,\,b}^{\quad a}\eta _{c}^{b}e^{\nu c}
\]
transforms to
\[
e_{d}^{\mu }M_{a}^{-1d}M_{e}^{a}R_{\mu \nu \,\,f}^{\quad e}M_{b}^{-1f}\eta
_{c}^{b}\left( M_{c}^{-1l}\right) ^{\dagger }e^{\nu l}
\]
and is thus $U(1,D-1)$ invariant. It is also Hermitian
\[
\left( e_{a}^{\mu }R_{\mu \nu \,\,b}^{\quad a}\eta _{c}^{b}e^{\nu c}\right)
^{\dagger }=-e_{c}^{\nu }\eta _{b}^{c}\eta _{e}^{b}R_{\mu \nu \,\,f}^{\quad
e}\eta _{a}^{f}e^{\mu a}=e_{a}^{\mu }R_{\mu \nu \,\,b}^{\quad a}\eta
_{c}^{b}e^{\nu c}
\]
The metric is defined by
\[
g_{\mu \nu }=\left( e_{\mu }^{a}\right) ^{\dagger }\eta _{b}^{a}e_{\nu }^{b}
\]
satisfy the property
\[
g_{\mu \nu }^{\dagger }=g_{\nu \mu }
\]
When the metric is decomposed into its real and imaginary parts:
\[
g_{\mu \nu }=G_{\mu \nu }+iB_{\mu \nu }
\]
the hermiticity property then implies the symmetries
\begin{eqnarray*}
G_{\mu \nu } &=&G_{\nu \mu } \\
B_{\mu \nu } &=&-B_{\nu \mu }
\end{eqnarray*}
The gauge invariant Hermitian action is given by
\[
I=\int d^{D}x\sqrt{G}e_{a}^{\mu }R_{\mu \nu \,\,b}^{\quad a}\eta
_{c}^{b}e^{\nu c}
\]
This action is analogous to the first order formulation of gravity obtained
by gauging the group $SO(1,D-1)$ One goes to the second order formalism by
integrating out the spin connection and substituting for it its value in
terms of the vielbein. The same structure is also present here and one can
solve for $\omega _{\mu \,\,b}^{\,a}$ in terms of the complex fields $e_{\mu
}^{a}$ resulting in an action that depends only on the fields $g_{\mu \nu }.$
It is worthwhile to stress that the above action, unlike others proposed to
describe nonsymmetric gravity \cite{M} is unique, except for the measure,
and unambiguous. Similar ideas have been proposed in the past based on
gauging the groups $O(D,D)$ \cite{MS} and $GL(D)$ \cite{Seigel}, in relation
to string duality, but the results obtained there are different from what is
presented here. The ordering of the terms in writing the action is done in a
way that generalizes to the noncommutative case.

The infinitesimal gauge transformations for $e_{\mu }^{a}$ is
\[
\delta e_{\mu }^{a}=\Lambda _{b}^{a}e_{\mu }^{b}
\]
which can be decomposed into real and imaginary parts by writing $e_{\mu
}^{a}=e_{0\mu }^{a}+ie_{1\mu }^{a},$ and $\Lambda _{b}^{a}=\Lambda
_{0b}^{a}+i\Lambda _{1b}^{a}$ to give
\begin{eqnarray*}
\delta e_{0\mu }^{a} &=&\Lambda _{0b}^{a}e_{0\mu }^{b}-\Lambda
_{1b}^{a}e_{1\mu }^{b} \\
\delta e_{1\mu }^{a} &=&\Lambda _{1b}^{a}e_{0\mu }^{b}+\Lambda
_{0b}^{a}e_{1\mu }^{b}
\end{eqnarray*}
The gauge parameters satisfy the constraints $\left( \Lambda _{b}^{a}\right)
^{\dagger }=-\eta _{c}^{b}\Lambda _{d}^{c}\eta _{a}^{d}$ which implies the
two constraints
\begin{eqnarray*}
\left( \Lambda _{0b}^{a}\right) ^{T} &=&-\eta _{c}^{b}\Lambda _{0d}^{c}\eta
_{a}^{d} \\
\left( \Lambda _{1b}^{a}\right) ^{T} &=&\eta _{c}^{b}\Lambda _{1d}^{c}\eta
_{a}^{d}
\end{eqnarray*}
From the gauge transformations of $e_{0\mu }^{a}$ and $e_{1\mu }^{a}$ one
can easily show that the gauge parameters $\Lambda _{0b}^{a}$ and $\Lambda
_{1b}^{a}$ can be chosen to make $e_{0\mu a}$ symmetric in $\mu $ and $a$
and $e_{1\mu a}$ antisymmetric in $\mu $ and $a$. This is equivalent to the
statement that the Lagrangian should be completely expressible in terms of $%
G_{\mu \nu }$ and $B_{\mu \nu }$ only, after eliminating $\omega _{\mu
\,\,b}^{\,a}$ through its equations of motion. In reality we have
\begin{eqnarray*}
G_{\mu \nu } &=&e_{0\mu }^{a}e_{0\nu }^{b}\eta _{ab}+e_{1\mu }^{a}e_{1\nu
}^{b}\eta _{ab} \\
B_{\mu \nu } &=&e_{0\mu }^{a}e_{1\nu }^{b}\eta _{ab}-e_{1\mu }^{a}e_{0\nu
}^{b}\eta _{ab}
\end{eqnarray*}
In this special gauge, where we define $g_{0\mu \nu }=e_{0\mu }^{a}e_{0\nu
}^{b}\eta _{ab}$ , $g_{0\mu \nu }g_{0}^{\nu \lambda }=\delta _{\mu
}^{\lambda },$ and use $e_{0\mu }^{a}$ to raise and lower indices we get
\begin{eqnarray*}
B_{\mu \nu } &=&-2e_{1\mu \nu } \\
G_{\mu \nu } &=&g_{0\mu \nu }-\frac{1}{4}B_{\mu \kappa }B_{\lambda \nu
}g_{0}^{\kappa \lambda }
\end{eqnarray*}
The last formula appears in the metric of the effective action in open
string theory \cite{CLNY}.

\section{Second Order Formulation}

We can express the Lagrangian in terms of $e_{\mu }^{a}$ only by solving the
$\omega _{\mu \,\,b}^{\,a}$ equations of motion
\begin{eqnarray*}
e_{a}^{\mu }e^{\nu b}\omega _{\nu \,\,b}^{\,c}+e_{b}^{\nu }e^{\mu c}\omega
_{\nu \,\,a}^{\,b}-e^{\mu b}e_{a}^{\nu }\omega _{\nu \,\,b}^{\,c}-e_{b}^{\mu
}e^{\nu c}\omega _{\nu \,\,a}^{\,b} &=& \\
\frac{1}{\sqrt{G}}\partial _{\nu }\left( \sqrt{G}\left( e_{a}^{\nu }e^{\mu
c}-e_{a}^{\mu }e^{\nu c}\right) \right) &\equiv &X_{\quad a}^{\mu c}
\end{eqnarray*}
where $X_{\quad a}^{\mu c}$ satisfy $\left( X_{\quad a}^{\mu c}\right)
^{\dagger }=-X_{\quad c}^{\mu a}.$ One has to be very careful in working
with a nonsymmetric metric
\begin{eqnarray*}
g_{\mu \nu } &=&e_{\mu }^{a}e_{\nu a} \\
g^{\mu \nu } &=&e^{\mu a}e_{\nu a} \\
g_{\mu \nu }g^{\nu \rho } &=&\delta _{\mu }^{\rho }
\end{eqnarray*}
but $g_{\mu \nu }g^{\mu \rho }\neq \delta _{\mu }^{\rho }.$ Care also should
be taken when raising and lowering indices with the metric.

Before solving the $\omega $ equations, we point out that the trace part of $%
\omega _{\mu \,\,b}^{\,a}$ (corresponding to the $U(1)$ part in $U(D)$) must
decouple from the other gauge fields. It is thus undetermined and decouples
from the Lagrangian after substituting its equation of motion. It imposes a
condition on the $e_{\mu }^{a}$%
\[
\frac{1}{\sqrt{G}}\partial _{\nu }\left( \sqrt{G}\left( e_{a}^{\nu }e^{\mu
a}-e_{a}^{\mu }e^{\nu a}\right) \right) \equiv X_{\quad a}^{\mu a}=0
\]
We can therefore assume, without any loss in generality, that $\omega _{\mu
\,\,b}^{\,a}$ is traceless $\left( \omega _{\mu \,\,a}^{\,a}=0\right) .$

Multiplying the $\omega -$equation with $e_{a}^{\kappa }e_{c}^{\rho }$ we
get
\[
\delta _{\kappa }^{\mu }\omega _{\nu \rho }^{\quad \nu }+\delta _{\rho
}^{\mu }\omega _{\nu \,\,\kappa }^{\,\,\nu }-\omega _{\kappa \rho }^{\quad
\mu }-\omega _{\rho \,\,\,\kappa }^{\,\,\mu }=X_{\,\,\rho \kappa }^{\mu }
\]
where
\begin{eqnarray*}
\omega _{\mu \nu }^{\quad \rho } &=&e_{\nu a}e^{\rho b}\omega _{\mu
\,\,b}^{\,\,a} \\
X_{\,\,\rho \kappa }^{\mu } &=&e_{\rho c}e_{\kappa }^{a}X_{\quad a}^{\mu c}
\end{eqnarray*}
Contracting by first setting $\mu =\kappa $ then $\mu =\rho $ we get the two
equations
\begin{eqnarray*}
3\omega _{\nu \rho }^{\quad \nu }+\omega _{\nu \,\,\rho }^{\,\,\nu }
&=&X_{\,\,\,\rho \mu }^{\mu } \\
\omega _{\nu \rho }^{\quad \nu }+3\omega _{\nu \,\,\rho }^{\,\,\nu }
&=&X_{\,\,\,\mu \rho }^{\mu }
\end{eqnarray*}
These could be solved to give
\begin{eqnarray*}
\omega _{\nu \rho }^{\quad \nu } &=&\frac{1}{8}\left( 3X_{\,\,\rho \mu
}^{\mu }-X_{\,\,\mu \rho }^{\mu }\right)  \\
\omega _{\nu \,\,\rho }^{\,\,\nu } &=&\frac{1}{8}\left( -X_{\,\,\rho \mu
}^{\mu }+3X_{\,\,\mu \rho }^{\mu }\right)
\end{eqnarray*}
Substituting these back into the $\omega -$equation we get
\[
\omega _{\kappa \rho }^{\quad \mu }+\omega _{\rho \,\,\kappa }^{\,\,\mu }=%
\frac{1}{8}\delta _{\kappa }^{\mu }\left( 3X_{\,\,\rho \mu }^{\mu
}-X_{\,\,\mu \rho }^{\mu }\right) +\frac{1}{8}\delta _{\rho }^{\mu }\left(
-X_{\,\,\kappa \mu }^{\mu }+3X_{\,\,\mu \kappa }^{\mu }\right) -X_{\,\,\rho
\kappa }^{\mu }\equiv Y_{\,\,\rho \kappa }^{\mu }
\]
We can rewrite this equation after contracting with $e_{\mu c}e_{\sigma }^{c}
$ to get
\[
\omega _{\kappa \rho \sigma }+e_{a}^{\mu }e_{\mu c}e_{\sigma }^{c}\omega
_{\rho \,\,\kappa }^{\,\,\,a}=g_{\sigma \mu }Y_{\,\,\rho \kappa }^{\mu
}\equiv Y_{\sigma \rho \kappa }
\]
By writing $\omega _{\rho \,\,\kappa }^{\,\,\,a}=\omega _{\rho \nu \kappa
}e^{\nu a}$ we finally get
\[
\left( \delta _{\kappa }^{\alpha }\delta _{\rho }^{\beta }\delta _{\sigma
}^{\gamma }+g^{\beta \mu }g_{\sigma \mu }\delta _{\rho }^{\alpha }\delta
_{\kappa }^{\gamma }\right) \omega _{\alpha \beta \gamma }=Y_{\sigma \rho
\kappa }
\]
To solve this equation we have to invert the tensor
\[
M_{\kappa \rho \sigma }^{\alpha \beta \gamma }=\delta _{\kappa }^{\alpha
}\delta _{\rho }^{\beta }\delta _{\sigma }^{\gamma }+g^{\beta \mu }g_{\sigma
\mu }\delta _{\rho }^{\alpha }\delta _{\kappa }^{\gamma }
\]
In the conventional case when all fields are real, the metric $g_{\mu \nu }$
is symmetric and $g^{\beta \mu }g_{\sigma \mu }=\delta _{\sigma }^{\beta }$
so that the inverse of $M_{\kappa \rho \sigma }^{\alpha \beta \gamma }$ is
simple. In the present case, because of the nonsymmetry of $g_{\mu \nu }$
this is fairly complicated and could only be solved by a perturbative
expansion. Writing $g_{\mu \nu }=G_{\mu \nu }+iB_{\mu \nu }$ and from the
definition $g^{\mu \nu }g_{\nu \rho }=\delta _{\mu }^{\rho }$ we get
\[
g^{\mu \nu }=a^{\mu \nu }+ib^{\mu \nu }
\]
where
\begin{eqnarray*}
a^{\mu \nu } &=&\left( G_{\mu \nu }+B_{\mu \kappa }G^{\kappa \lambda
}B_{\lambda \nu }\right) ^{-1} \\
&=&G^{\mu \nu }-G^{\mu \kappa }B_{\kappa \lambda }G^{\lambda \sigma
}B_{\sigma \eta }G^{\eta \nu }+O(B^{4})
\end{eqnarray*}
\[
b^{\mu \nu }=-G^{\mu \kappa }B_{\kappa \lambda }G^{\lambda \nu }+G^{\mu
\kappa }B_{\kappa \lambda }G^{\lambda \sigma }B_{\sigma \tau }G^{\tau \rho
}B_{\rho \eta }G^{\eta \nu }+O(B^{5})
\]
We have defined $G^{\mu \nu }G_{\nu \rho }=\delta _{\rho }^{\mu }$. This
implies that
\begin{eqnarray*}
g^{\mu \alpha }g_{\nu \alpha } &\equiv &\delta _{\nu }^{\mu }+L_{\nu }^{\mu }
\\
L_{\nu }^{\mu } &=&iG^{\mu \rho }B_{\rho \nu }-2G^{\mu \rho }B_{\rho \sigma
}G^{\sigma \alpha }B_{\alpha \nu }+O(B^{3})
\end{eqnarray*}
The inverse of $M_{\kappa \rho \sigma }^{\alpha \beta \gamma }$ defined by
\[
N_{\alpha \beta \gamma }^{\sigma \rho \kappa }M_{\sigma \rho \kappa
}^{\alpha ^{\prime }\beta ^{\prime }\gamma ^{\prime }}=\delta _{\alpha
}^{\alpha ^{\prime }}\delta _{\beta }^{\beta ^{\prime }}\delta _{\gamma
}^{\gamma ^{\prime }}
\]
is evaluated to give
\begin{eqnarray*}
N_{\alpha \beta \gamma }^{\sigma \rho \kappa } &=&\frac{1}{2}\left( \delta
_{\gamma }^{\sigma }\delta _{\beta }^{\rho }\delta _{\alpha }^{\kappa
}+\delta _{\beta }^{\sigma }\delta _{\alpha }^{\rho }\delta _{\gamma
}^{\kappa }-\delta _{\alpha }^{\sigma }\delta _{\gamma }^{\rho }\delta
_{\beta }^{\kappa }\right)  \\
&&-\frac{1}{4}\left( \delta _{\beta }^{\kappa }\delta _{\alpha }^{\sigma
}L_{\gamma }^{\rho }+\delta _{\alpha }^{\kappa }\delta _{\gamma }^{\sigma
}L_{\beta }^{\rho }-\delta _{\gamma }^{\kappa }\delta _{\beta }^{\sigma
}L_{\alpha }^{\rho }\right)  \\
&&+\frac{1}{4}\left( L_{\gamma }^{\kappa }\delta _{\beta }^{\sigma }\delta
_{\alpha }^{\rho }+L_{\beta }^{\kappa }\delta _{\alpha }^{\sigma }\delta
_{\gamma }^{\rho }-L_{\alpha }^{\kappa }\delta _{\gamma }^{\sigma }\delta
_{\beta }^{\rho }\right)  \\
&&-\frac{1}{4}\left( \delta _{\alpha }^{\kappa }L_{\gamma }^{\sigma }\delta
_{\beta }^{\rho }+\delta _{\gamma }^{\kappa }L_{\beta }^{\sigma }\delta
_{\alpha }^{\rho }-\delta _{\beta }^{\kappa }L_{\alpha }^{\sigma }\delta
_{\gamma }^{\rho }\right) +O(L^{2})
\end{eqnarray*}
This enables us to write
\[
\omega _{\alpha \beta \gamma }=N_{\alpha \beta \gamma }^{\sigma \rho \kappa
}Y_{\rho \sigma \kappa }
\]
and finally
\[
\omega _{\mu \,\,b}^{\,a}=e^{\beta a}e_{b}^{\gamma }\omega _{\mu \beta
\gamma }
\]
It is clear that the leading term reproduces the Einstein-Hilbert action
plus contributions proportional to $B_{\mu \nu }$ and higher order terms.
The most difficult task is to show that the Lagrangian is completely
expressible in terms of $G_{\mu \nu }$ and $B_{\mu \nu }$ only. The other
components of $e_{0\mu }^{a}$ and $e_{1\mu }^{a}$ should disappear. We have
argued from the view point of gauge invariance that this must happen, but it
will be nice to verify this explicitly, to leading orders. We can check that
in the flat approximation for gravity with $G_{\mu \nu }$ taken to be $%
\delta _{\mu \nu }$, the $B_{\mu \nu }$ field gets the correct kinetic
terms. First we write
\begin{eqnarray*}
e_{\mu }^{a} &=&\delta _{\mu }^{a}-\frac{i}{2}B_{\mu a} \\
e_{\mu a} &=&\delta _{\mu }^{a}+\frac{i}{2}B_{\mu a}
\end{eqnarray*}
and the inverses
\begin{eqnarray*}
e^{\mu a} &=&\delta _{\mu }^{a}-\frac{i}{2}B_{\mu a} \\
e_{a}^{\mu } &=&\delta _{\mu }^{a}+\frac{i}{2}B_{\mu a}
\end{eqnarray*}
The $\omega _{\mu \,\,a}^{\,\,a}$equation implies the constraint
\[
X_{\,\quad a}^{\mu a}=\partial _{\nu }\left( e_{a}^{\mu }e^{\nu
a}-e_{a}^{\nu }e^{\mu a}\right) =0
\]
This gives the gauge fixing condition
\[
\partial ^{\nu }B_{\mu \nu }=0
\]
We then evaluate
\[
X_{\,\,\rho \kappa }^{\mu }=-\frac{i}{2}\left( \partial _{\rho }B_{\kappa
\mu }+\partial _{\kappa }B_{\rho \mu }\right)
\]
This together with the gauge condition on $B_{\mu \nu }$ gives
\[
Y_{\,\,\rho \kappa }^{\mu }=\frac{i}{2}\left( \partial _{\rho }B_{\kappa \mu
}+\partial _{\kappa }B_{\rho \mu }\right)
\]
and finally
\[
\omega _{\mu \nu \rho }=-\frac{i}{2}\left( \partial _{\mu }B_{\nu \rho
}+\partial _{\nu }B_{\mu \rho }\right)
\]
When the $\omega _{\mu \nu \rho }$ is substituted back into the Lagrangian,
and after integration by parts one gets
\begin{eqnarray*}
L &=&\omega _{\mu \nu \rho }\omega ^{\nu \rho \mu }-\omega _{\mu
\,\,}^{\,\,\,\mu \rho }\omega _{\nu \rho }^{\quad \nu } \\
&=&-\frac{1}{4}B_{\mu \nu }\partial ^{2}B^{\mu \nu }
\end{eqnarray*}
This is identical to the usual expression
\[
\frac{1}{12}H_{\mu \nu \rho }H^{\mu \nu \rho }
\]
where
\[
H_{\mu \nu \rho }=\partial _{\mu }B_{\nu \rho }+\partial _{\nu }B_{\rho \mu
}+\partial _{\rho }B_{\mu \nu }
\]
We have therefore shown that in $D$ dimensions one must start with $2D^{2}$
real components $e_{\mu }^{a}$, subject to gauge transformations with $D^{2}$
real parameters. The resulting Lagrangian depends on $D^{2}$ fields, with $%
\frac{D\left( D+1\right) }{2}$ symmetric components $G_{\mu \nu }$ and $%
\frac{D\left( D-1\right) }{2}$ antisymmetric components $B_{\mu \nu }.$

\section{Noncommutative Gravity}

At this stage, and having shown that it is perfectly legitimate to formulate
a theory of gravity with nonsymmetric complex metric, based on the idea of
gauge invariance of the group $U(1,D-1).$ It is not difficult to generalize
the steps that led us to the action for complex gravity to spaces where
coordinates do not commute, or equivalently, where the usual products are
replaced with star products.

First the gauge fields are subject to the gauge transformations
\[
\widetilde{\omega }_{\mu \,\,b}^{\,a}=M_{c}^{a}\ast \omega _{\mu
\,\,d}^{\,c}\ast M_{\ast b}^{-1d}-M_{c}^{a}\ast \partial _{\mu }M_{\ast
b}^{-1c}
\]
where $M_{\ast a}^{-1b}$ is the inverse of $M_{b}^{a}$ with respect to the
star product. The curvature is now
\[
R_{\mu \nu \,\,b}^{\quad a}=\partial _{\mu }\omega _{\nu
\,\,b}^{\,a}-\partial _{\nu }\omega _{\mu \,\,b}^{\,a}+\omega _{\mu
\,\,c}^{\,a}\ast \omega _{\nu \,\,b}^{\,c}-\omega _{\nu \,\,c}^{\,a}\ast
\omega _{\mu \,\,b}^{\,c}
\]
which transforms according to
\[
\widetilde{R}_{\mu \nu \,\,b}^{\quad a}=M_{c}^{a}\ast R_{\mu \nu
\,\,d}^{\quad c}\ast M_{\ast b}^{-1d}
\]
Next we introduce the vielbeins $e_{\mu }^{a}$ and their inverse defined by
\begin{eqnarray*}
e_{\ast a}^{\nu }\ast e_{\mu }^{a} &=&\delta _{\mu }^{\nu } \\
e_{\nu }^{a}\ast e_{\ast b}^{\nu } &=&\delta _{b}^{a}
\end{eqnarray*}
which transform to
\begin{eqnarray*}
\widetilde{e}_{\mu }^{a} &=&M_{b}^{a}\ast e_{\mu }^{b} \\
\widetilde{e}_{\ast a}^{\mu } &=&\widetilde{e}_{b}^{\mu }\ast M_{\ast
a}^{-1b}
\end{eqnarray*}
The complex conjugates for the vielbeins are defined by
\begin{eqnarray*}
e_{\mu a} &\equiv &\left( e_{\mu }^{a}\right) ^{\dagger } \\
e_{\ast }^{\mu a} &\equiv &\left( e_{\ast a}^{\mu }\right) ^{\dagger }
\end{eqnarray*}
Finally we define the metric
\[
g_{\mu \nu }=\left( e_{\mu }^{a}\right) ^{\dagger }\eta _{b}^{a}\ast e_{\nu
}^{b}
\]
The $U(1,D-1)$ gauge invariant Hermitian action is
\[
I=\int d^{D}x\sqrt{G}\left( e_{\ast a}^{\mu }\ast R_{\mu \nu \,\,b}^{\quad
a}\eta _{c}^{b}\ast e_{\ast }^{\nu c}\right)
\]
This action differs from the one considered in the commutative case by
higher derivatives terms proportional to $\theta ^{\mu \nu }$. It would be
very interesting to see whether these terms could be reabsorbed by
redefining the field $B_{\mu \nu }$, or whether the Lagrangian reduces to a
function of $G_{\mu \nu }$ and $B_{\mu \nu }$ and their derivatives only.

The connection of this action to the gravity action derived for
noncommutative spaces based on spectral triples (\cite{CFF},\cite{CGF},\cite
{C}) remains to be made. In order to do this one must understand the
structure of Dirac operators for spaces with deformed star products.

\section{Conclusions}

We have shown that it is possible to combine the tensors $G_{\mu \nu }$ and $%
B_{\mu \nu }$ into a complexified theory of gravity in $D$ dimensions by
gauging the group $U(1,D-1)$. The Hermitian gauge invariant action is a
direct generalization of the first order formulation of gravity obtained by
gauging the Lorentz group $SO(1,D-1)$. The Lagrangian obtained is a function
of the complex fields $e_{\mu }^{a}$ and reduces to a function of $G_{\mu
\nu }$ and $B_{\mu \nu }$ only. This action is generalizable to
noncommutative spaces where coordinates do not commute, or equivalently,
where the usual products are deformed to star products. It is remarkable
that the presence of a constant background field in open string theory
implies that the metric of the target space becomes nonsymmetric and that
the tangent manifold for space-time does not have only the Lorentz symmetry
but the larger $U(1,D-1)$ symmetry. The results shown here, can be improved
by computing the second order action to include higher order terms in the $%
B_{\mu \nu }$ expansion and to see if this can be put in a compact form.
Similarly the computation has to be repeated in the noncommutative case to
see whether the $\theta ^{\mu \nu }$ contributions could be simplified. It
is also important to determine a link between this formulation of
noncommutative gravity and the Connes formulation based on the
noncommutative geometry of spectral triples. To make such connection many
points have to be clarified, especially the structure of the Dirac operator
for such a space. This and other points will be explored in future
publication

\end{document}